\documentclass{llncs}

\usepackage{amsmath}
\usepackage{amssymb}
\usepackage{graphicx}
\usepackage{relsize}
\usepackage{url}

\newcommand{\partd}[2]{\frac{\partial {#1}}{\partial {#2}}}

\newcommand{\bx}{{\bf x}}
\newcommand{\by}{{\bf y}}

\newcommand{\bw}{{\bf w}}
\newcommand{\ba}{{\bf a}}
\newcommand{\bb}{{\bf b}}

\newcommand{\avec}[1]{\left\langle{#1}\right\rangle}
\newcommand{\conj}{\,\mathrel{{\mathlarger{\wedge}}}\,}
\newcommand{\disj}{\,\mathrel{\mathlarger{\vee}}\,}

\newcommand{\R}{\mathbb{R}}
\newcommand{\eq}{\Longleftrightarrow}
\newcommand{\Abs}[1]{\left|#1\right|}
\newcommand{\abs}[1]{|#1|}
\newcommand{\dih}{\mathrel{\rm dih}}

\begin{document}

\title{Formal Verification of Nonlinear Inequalities with Taylor Interval Approximations}

\author{Alexey Solovyev \and Thomas C. Hales\thanks{Research supported by NSF grant 0804189 and a grant from the Benter Foundation.}}

\institute{Department of Mathematics, University of Pittsburgh,\\
Pittsburgh, PA 15260, USA}

\maketitle
\begin{abstract}
We present a formal tool for verification of multivariate nonlinear inequalities. Our verification method is based on interval arithmetic with Taylor approximations. Our tool is implemented in the HOL Light proof assistant and it is capable to verify multivariate nonlinear polynomial and non-polynomial inequalities on rectangular domains. One of the main features of our work is an efficient implementation of the verification procedure which can prove non-trivial high-dimensional inequalities in several seconds. We developed the verification tool as a part of the Flyspeck project (a formal proof of the Kepler conjecture). The Flyspeck project includes about 1000 nonlinear inequalities. We successfully tested our method on more than 100 Flyspeck inequalities and estimated that the formal verification procedure is about 3000 times slower than an informal verification method implemented in C++. We also describe future work and prospective optimizations for our method.
\end{abstract}

\section{Introduction}
In this paper, we present a tool for formal verification of nonlinear inequalities in HOL Light~\cite{HOLL}. Our tool can verify multivariate polynomial and non-polynomial inequalities on rectangular domains. The verification technique is based on interval arithmetic with Taylor approximations. A short user manual describing our tool is available~\cite{nonlinear-manual}. Solovyev's thesis~\cite{solovyev-thesis} contains additional information about the verification tool and the corresponding formal techniques.

Our work is an integral part of the Flyspeck project~\cite{hales:DSP:2006:432,website:FlyspeckProject}. This project was launched in 2003 by T.~Hales to produce a complete formal verification of Hales' proof of the Kepler conjecture~\cite{Hales:2006:DCG,DSP}. There are several major computationally extensive verification problems in the Flyspeck project. One of these problems is a formval verification of about 1000 multivariate nonlinear inequalities. We have successfully tested our formal verification tool on several simple Flyspeck nonlinear inequalities (we have verified 130 inequalities). In theory, almost all Flyspeck inequalities can be verified with our formal verification procedure. A rough estimate shows that the current formal procedure is about 3000 times slower than the corresponding informal verification algorithm in C++~\cite{hales-algorithm}. With this estimate, it will take more than 4 years to verify all Flyspeck nonlinear inequalities formally on a single computer (the informal procedure requires about 9 hours).

There exist other formal methods for verification of nonlinear inequalities. First of all, general quantifier elimination procedures may be used to solve some polynomial inequalities~\cite{tarski-decision,collins,mclaughlin-harrison}. Another method for proving polynomial inequalities is known as sums-of-squares (SOS) method~\cite{harrison-sos}.

A tool called MetiTarski~\cite{metitarski-prover,metitarski-future} is capable to verify multivariate polynomial and non-polynomial inequalities on unbounded domains. It approximates non-polynomial functions by suitable polynomial bounds and then applies quantifier elimination procedures for resulting polynomials.

The Bernstein polynomial technique~\cite{roland-thesis} allows to verify multivariate polynomial inequalities. Each polynomial can be written as a sum of polynomials in the Bernstein polynomial basis. Coefficients of this representation give bounds of the polynomial itself. A complete formal implementation of this method is done in PVS~\cite{MN12}. Non-polynomial inequalities must be first converted into polynomial inequalities by finding polynomial bounds. One way to find polynomial bounds is to use Taylor model approximations~\cite{roland-taylor}. R.~Zumkeller's thesis describes this method in details~\cite{roland-thesis}. He also implemented an informal global optimization tool based on Bernstein polynomials~\cite{website:sergei} in Haskell.

There exists a tool in the PVS proof assistant which uses the same technique as our tool (interval arithmetic with Taylor approximations)~\cite{DLM09} but this tool works only with univariate functions.

Methods based on quantifier elimination procedures do not scale well when the number of variables grows and when inequalities become more complicated. The Bernstein polynomial technique works well for polynomial inequalities but does not show very good results for inequalities involving special functions in high dimensions.

\section{Verification of Nonlinear Inequalities}
\subsection{Nonlinear Inequalities and Interval Taylor Approximations}

Consider the problem: prove that 
\[
\forall {\bf x} \in \R^n, {\bf x} \in D \implies f({\bf x}) < 0.
\]
$D$ is assumed to be a rectangle given by $D = \{(x_1, \ldots, x_n)\ |\ a_i \le x_i \le b_i\} = [{\bf a}, {\bf b}]$. We also assume that $f({\bf x})$ is twice continuously differentiable in an open domain $U \supset D$.

One way to solve the problem is to consider a finite partition of $D = \bigcup_j D^{j}$ such that each $D^{j}$ is rectangular. Also, we assume that $\bar{f}(D^{j}) < 0$ where $\bar{f}$ is an interval approximation of $f$ (that is, $\bar{f}(D^{j})$ is the interval corresponding to the interval evaluation of $f(x_1,\ldots,x_n)$ for input intervals $x_i \in [a_i^{j}, b_i^{j}]$; clearly, $\bar{f}(D) < 0 \implies f(D) < 0$). It is easy to see that such a partition always exists if $f$ is continuous, $f(D) < 0$, and $f$ can be arbitrary well approximated by $\bar{f}$ on sufficiently small domains. (It follows by the compactness argument: for each point $x \in D$ there is a small rectangle $D^{j}$ such that $x \in \rm{interior}(D^{j})$ and $\bar{f}(D^{j}) < 0$; $D$ is compact, so there are finitely many rectangles $D^{j}$ such that $D = \bigcup_j D^{j}$.)

The main difficulty is finding a suitable partition $\{D^{j}\}$. The easiest way is the following. Let $D^{0} = D$ and compute $\bar{f}(D^{0})$. If this value is less than $0$ (in the interval sense), then we are done. Otherwise divide $D^{0}$ into two regions $D^{0} = D^{1}_1 \cup D^{1}_2$. Then repeat the procedure for regions with upper index $1$. In general, either $\bar{f}(D^k_j) < 0$ or we get $D^k_j = D^{k+1}_{2j-1} \cup D^{k+1}_{2j}$. If we divide each region such that sizes of new regions become arbitrary small in all dimensions, then the process will eventually stop and a suitable partition of $D$ will be found. An easy way to achieve this goal is to divide each region in half along the coordinate for which its size is maximal, i.e., if $D^{k}_j = \{a_i \le x_i \le b_i\} = [{\bf a}, {\bf b}]$ and $b_m - a_m = \max_i \{b_i - a_i\}$, then set $D^{(k+1)_{2j-1}} = [{\bf a}, {\bf b}^{(m,y)}]$ and $D^{(k+1)}_{2j} = [{\bf a}^{(m,y))}, {\bf b}]$. Here, $y = (a_m + b_m) / 2$ and ${\bf a}^{(m,y)}$ equals to ${\bf a}$ with the $m$-th component replaced by~$y$.

As the result of the procedure above, we get a finite set of subregions $S = \{D^k_i\}$ with the property: for each $D^k_i \in S$ either $\bar{f}(D^k_i) < 0$ or $D^k_i = D^{k+1}_{i_1} \cup D^{k+1}_{i_2}$. In the last case, the verification relies on a trivial theorem
\[
D = D_1 \cup D_2 \conj f(D_1) < 0 \conj f(D_2) < 0 \implies f(D) < 0.
\]

Interval arithmetic works for any continuous function (at least in theory where numerical errors are not considered) but it is not very efficient in general. This is due to the dependency problem when even a simple function could require a lot of subdivisions in order to get the result on the full domain. Even a trivial inequality $f(x) = x - x < 1$ will require subdivisions for the domain $x \in [0,1]$. Indeed, $\bar{f}([0,1]) = [0,1] - [0,1] = [-1,1]$. Of course, we can simplify $x - x = 0$ but it is not possible to do for a function $f(x) = x - \arctan(x)$ which has similar behaviour near $0$. For this function, $\bar{f}([0,1]) = [0,1] - [0, \pi/4] = [-\pi/4, 1]$ and we don't get $f(x) < 1$. One way to decrease the dependency problem is to use Taylor approximations for computing bounds of $f$ on a given domain $D$.

Fix ${\bf y} \in D = [{\bf a}, {\bf b}]$, then we can write
\[
f({\bf x}) = f({\bf y}) + \sum_{i = 1}^n \partd{f}{x_i}({\bf y}) (y_i - x_i) +
\frac{1}{2}\sum_{i, j = 1}^n \partd{^2 f}{x_i \partial x_j}({\bf p}) (y_i - x_i) (y_j - x_j)
\]
where ${\bf p} \in [{\bf a}, {\bf b}]$. Let ${\bf w} = \max\{{\bf y} - {\bf a}, {\bf b} - {\bf y}\}$ (all operations are componentwise). Suppose we have interval bounds for $f({\bf y}) \in [f_0^{lo}, f_0^{hi}]$, $\partd{f}{x_i}({\bf y}) \in [f_i^{lo}, f_i^{hi}]$ and $\partd{^2 f}{x_i \partial x_j}({\bf t}) \in [f_{ij}^{lo}, f_{ij}^{hi}]$ for all ${\bf t} \in D$. We can write
\begin{multline*}
\forall {\bf x} \in D,\ f({\bf x}) \le f({\bf y}) + \sum_{i = 1}^n \Abs{\partd{f}{x_i}({\bf y})} w_i +
\frac{1}{2}\sum_{i, j = 1}^n \Abs{\partd{^2 f}{x_i \partial x_j}(\xi)} w_i w_j \\
\le f_0^{hi} + \sum_{i = 1}^n \Abs{[f_i^{lo}, f_i^{hi}]} w_i +
\frac{1}{2} \sum_{i, j = 1}^n \Abs{[f_{ij}^{lo}, f_{ij}^{hi}]} w_i w_j.
\end{multline*}
Absolute values of intervals are defined by $\Abs{[a, b]} = \max\{-a, b\}$.

Let's see how well this approximation works on examples. Again, take $f(x) = x - x$ and $D = [0,1]$. We compute $f'(x) = 1 - 1 = 0$ and $f''(x) = 0$. Set $y = 0.5$ and $w = 0.5$. Suppose $\bar{f}(0.5) = [0.4, 0.6] - [0.4, 0.6] = [-0.2, 0.2]$ (we deliberately take a very poor interval approximation), then
\[
\forall x \in [0,1],\ f(x) \le \bar{f}(0.5)^u + \sum_{i=1}^1 0 \times 0.5 + 
\sum_{i, j = 1}^1 0 \times 0.5 \times 0.5 = 0.2 < 1.
\]
In the same way, for $f(x) = x - \arctan x$ we get $f'(x) = 1 - \frac{1}{1 + x^2}$, $f''(x) = \frac{-2x}{(1 + x^2)^2}$. If $x \in [0,1]$, then $f''(x) \in [-2, 0] = [f_{11}^{lo}, f_{11}^{hi}]$ and hence $\abs{f''(x)} \le 2$. We compute
\[
\forall x \in [0,1],\ f(x) \le 0.04 + 0.21 \times 0.5 + 2 \times 0.5^3 \le 0.4.
\]

We see that interval arithmetic with Taylor approximations works much better. Moreover, we don't need to abandon direct interval approximations completely: every time when we have to verify whether $f(D_i) < 0$ we can first find an interval approximation $\bar{f}(D_i)$  and then compute a Taylor approximation. If we don't get the inequality in both cases, then we subdivide the domain.

%We will formally define Taylor interval approximations as interval bounds of $f$ and its first-order partial derivatives at a fixed point ${\bf y} \in D$ and interval bounds of second-order partial derivatives for all ${\bf t} \in D$. Moreover, our formal definition will also include domain bounds ${\bf a}, {\bf b}$, the point ${\bf y}$ and the width parameters ${\bf w}$. See the end of Section~\ref{formal-theories} for the formal definition.

One simple trick which can be done with both interval and Taylor interval approximations is estimation of partial derivatives on a given domain. If it happens that $f_j(D_k) = \partd{f}{x_j}(D_k) \le 0$ or $f_j(D_k) \ge 0$ then it will be immediately possible to restrict further verifications to the boundary of $D_k = [{\bf a}, {\bf b}]$. Indeed, if $f_j(D_k) \le 0$ and $f(D_k|_{x_j = a_j}) < 0$ then $f(D_k) < 0$ since the function is decreasing along the $j$-th coordinate and its maximal value is attained at $x_j = a_j$. The same is true for increasing functions (consider $D_k|_{x_j = b_j}$). Moreover, if $\{x_j = a_j\}$ ($\{x_j = b_j\}$) is not on the boundary of the main domain $D_k$, then it is possible to completely ignore any further verifications for the region $D_k$. Indeed, if the restriction of $D_k$ is not on the boundary of the original domain, then there is another subdomain $D_j$ such that the restriction of $D_k$ is a subset of $D_j$ and the inequality is true on $D_j$. However, we need to be careful. Consider an example. Suppose $f(x) = -x^2 - 1$ and $D = [-1,1]$. Assume that we have $D_1 = [-1, 0]$ and $D_2 = [0,1]$. We get $f'(x) = -2x \ge 0$ on $[-1,0]$. Hence, the function is increasing and we can consider the restricted domain $\{0\}$ which is not on the boundary of $[-1,1]$. Also, $f'(x) = -2x \le 0$ on $[0,1]$ and we again get $\{0\}$ as the restriction of $[0,1]$. If we don't continue verifications in both cases, then we will not be able to verify the inequality. In order to avoid this problem, we always check a strict inequality for decreasing functions, that is, we test if $f_j({\bf x}) \ge 0$ or $f_j({\bf x}) < 0$.

Another trick is to check convexity of a function before subdividing a domain $D_k$. If we need to subdivide $D_k$ and find that $f_{jj}(D) = \partd{^2 f}{x_j \partial x_j}(D) \ge 0$, then it is enough to verify $f(D_k|_{x_j = a_j}) < 0$ and $f(D_k|_{x_j = b_j}) < 0$. By convexity of $f$ (i.e., $f$ attains its maximum on the boundary), we get $f(D_k) < 0$ from these two inequalities.

\subsection{Solution Certificate Search Procedure}
\label{nonlinear-certificate}

An informal verification procedure based on the ideas presented above has been developed in C++ for informal verification of Flyspeck nonlinear inequalities~\cite{hales-algorithm}. The starting point of our implementation of a formal procedure for verification of nonlinear inequalities is a port of this original C++ program into OCaml. This OCaml program informally verifies a given nonlinear inequality on a rectangular domain by finding Taylor interval approximations and subdividing domains if necessary. The result of this program is just a boolean value: yes or no, the inequality true or false (there is the third option: verification could fail due to numerical instability or when subdomains become very small without any definite results).

We have modified the OCaml informal verification procedure such that it returns a partition of the original domain in a special tree-like structure which also contains all necessary information about verification steps for each subdomain. We call this structure a solution certificate for a given nonlinear inequality. The informal procedure is called the solution certificate search procedure.

A solution certificate is defined with the following OCaml record
\begin{verbatim}
type result_tree =
  | Result_false
  | Result_pass
  | Result_mono of mono_status list * result_tree
  | Result_glue of (int * bool * result_tree * result_tree)
  | Result_pass_mono of mono_status
  | Result_pass_ref of int
\end{verbatim}
The record \verb'mono_status' contains monotonicity information (i.e., whether some first-order partial derivative is negative or positive).

A simplified solution certificate search algorithm is given below in OCaml-like pseudo code.
\begin{verbatim}
let search f dom =
  let taylor_inteval = {find Taylor approximation of f on dom}
  let bounds = {taylor_interval bounds}
  if bounds >= 0 then
    Result_false
  else if bounds < 0 then
    Result_pass
  else
    let d_bounds = {find bounds of partial derivatives}
    let mono = {list of negative and positive partial derivatives}
    if {mono is not empty} then
       let r_dom = {restrict dom using information from mono}
         Result_mono mono (search f r_dom)
    else
      let dd_bounds = {find bounds of second partial derivatives}
      if {the j-th second partial derivative is non-negative} then
        let dom1, dom2 = {restrict dom along j}
        let c1 = search f dom1
        let c2 = search f dom2
          Result_glue (j, true, c1, c2)
      else
        let j = {find j such that b_i - a_i is maximal}
        let dom1, dom2 = {split dom along j}
        let c1 = search f dom1
        let c2 = search f dom2
          Result_glue (j, false, c1, c2)
\end{verbatim}
If the inequality $f(x) < 0$ holds on $D$, then the algorithm (applied to $f$ and $D$) will return a solution certificate which does not contain \verb'Result_false' nodes (of course, the real algorithm could fail due to numerical instabilities and rounding errors). A solution certificate does not contain any explicit information about subdomains for which verification must be performed. All subdomains can be restored from a solution certificate and the initial domain $D$. For each \verb'Result_glue(j, false, c1, c2)' node, it is necessary to split the domain in two halves along the $j$-th coordinate. The second argument is the convexity flag. If it is true, then the current domain must be restricted to its left and right boundaries along the $j$-th coordinate. For new subdomains, the node contains their solution certificates: \verb'c1' and \verb'c2'. The domain also has to be modified for \verb'Result_mono' nodes. Each node of this type contains a list of indices and boolean parameters (packed in \verb'mono_status' record) which indicate for which partial derivatives the monotonicity argument should be applied; boolean parameters determine if the corresponding partial derivatives are positive or negative.

The simplified algorithm never returns nodes of type \verb'Result_pass_mono'. The real solution certificate search algorithm is a little more complicated. Every time when monotonicity argument is applied, it checks if the restricted domain is on the boundary of the original domain or not (the original domain is an argument of the algorithm). If the restricted domain is not on the boundary of the original domain, then \verb'Result_pass_mono' will be returned.

If a solution certificate contains nodes of type \verb'Result_pass_mono', then it is necessary to transform such a certificate to get new certificates which can be formally verified. Indeed, suppose we have a \verb'Result_pass_mono' node and the corresponding domain is $D_k$. \verb'Result_pass_mono' requires to apply the monotonicity argument to $D_k$, that is, to restrict this domain to its boundary along some coordinate. But it doesn't contain any information on how to verify the inequality on the restricted subdomain. We can only claim that there is another subdomain $D_j$ (corresponding to some other node of a solution certificate) such that the restriction of $D_k$ is a subset of $D_j$. In other words, to verify the inequality on $D_k$, we first need to find $D_j$ such that the restriction of $D_k$ is a subset of $D_j$ and such that the inequality can be verified on $D_j$. To solve this problem, we transform a given solution certificate into a list of solution certificates and subdomains for which these new solution certificates work. Each solution certificate in the list may refer to previous solution certificates with \verb'Result_ref'. The last solution certificate in the list corresponds to the original domain. The transformation algorithm is the following
\begin{verbatim}
let transform certificate acc =
   let sub_certs = {find all maximal sub-certificates 
                   which does not contain Result_pass_mono}
   if {sub_certs contains certificate} then 
      {add certificate to acc and return acc}
   else
      let sub_certs = {remove certificates consisting of single 
                       Result_ref from sub_certs}
      let paths = {find paths to sub-certificates in sub_cert}
      let _ = {add sub_certs and the corresponding paths to acc}
      let new_cert1 = {replace all sub_certs in certificate 
                       with references}
      let new_cert2 = {replace Result_pass_mono nodes in new_cert1 
                       if they can be verified using subdomains 
                       defined by paths in acc}
         transform new_cert2 acc
\end{verbatim}
This algorithm maintains a list \verb'acc' of solution certificates which do not contain nodes of type \verb'Result_pass_mono'. The list also contains paths to subdomains corresponding to certificates. Each path is a list of pairs and it can be used to construct the corresponding subdomain starting from the original domain. Each pair is one of \verb'("l", i)', \verb'("r", i)', \verb'("ml", i)', or \verb'("mr", i)' where $i$ is an index. \verb'"l"' and \verb'"r"' labels correspond to left and right subdomains after splitting. \verb'"ml"' and \verb'"mr"' correspond to left and right restricted subdomains. The index $i$ specifies the coordinate along which the operation must be performed. When a reference node \verb'Result_ref' is generated for a sub-certificate at the $j$-th position in the accumulator list \verb'acc', then the argument of \verb'Result_ref' is $j$. 

% Formal Theories
\section{Formal Verification}
\label{formal}
The first step of developing a formal verification procedure is formalization of all necessary theories involving the multivariate Taylor theorem and related topics. Standard HOL Light libraries contain a formalization of Euclidean vector space~\cite{harrison-euclidean} and define general Frechet derivatives and Jacobian matrices for working with first-order partial derivatives. Also, HOL Light contains the general univariate Taylor theorem. We formalized all other important results including the theory of partial derivatives, the equality of second-order mixed partial derivatives, the multivariate Taylor formula with the second-order error term.

The main formal verification step is to compute a formal Taylor interval approximation for a function $f:\R^n \to \R$ on a given domain $D = [\ba, \bb]$. Each formal Taylor approximation includes the following data: a point $\by = (\ba + \bb) / 2 \in D$, a vector $\bw$ which estimates the width of the domain and has the property $\bw \ge \max\{\bb - \by, \by - \ba\}$ (all operations are componentwise), an interval bound of $f(\by) \in [f^{lo}, f^{hi}]$, interval bounds of partial derivatives $f_i(\by) \in [f_i^{lo}, f_i^{hi}] = d_i$ for all $i = 1,\ldots,n$, interval bounds of second-order partial derivatives on the full domain $f_{ij}(\bx) \in [f_{ij}^{lo}, f_{ij}^{hi}] = d_{ij}$ for all $i = 1,\ldots,n$, $j \le i$, and $\bx \in D$. Based on this data, an interval approximation of $f(\bx)$ and its partial derivatives on $D$ can be computed. For instance, the following theorem gives an interval approximation of $f(\bx)$ when $n = 2$
\begin{align*}
w_1 \abs{d_1} + w_2 \abs{d_2} \le b
	&\conj w_1(w_1 \abs{d_{1,1}}) + w_2(w_2\abs{d_{2,2}} + 2 w_1 \abs{d_{2,1}}) \le e\\
&\conj b + 2^{-1} e \le a \conj l \le f^{lo} - a \conj f^{hi} + a \le h\\
&\implies \bigl(\forall \bx,\ \bx \in [\ba, \bb] \implies f(\bx) \in [l,h]\bigr).
\end{align*}
(Here, $\abs{d_i} = \abs{[f_i^{lo}, f_i^{hi}]} = \max\{-f_i^{lo}, f_i^{hi}\}$.)

Formal computations of Taylor interval approximations require a lot of basic arithmetic operations. We implemented efficient procedures for working with natural numbers and real numbers in HOL Light. Our implementation of formal natural number arithmetic works with numerals in an arbitrary fixed base. Our implementation improves the performance of standard HOL Light arithmetic operations with natural numbers by the factor $\log_2 b$ (where $b$ is a fixed base constant) for linear operations (in the size of input arguments) and by the factor $(\log_2 b)^2$ for quadratic operations. We approximate real numbers with floating-point numbers which have fixed precision of the mantissa. This precision is controlled by an informal parameter which specifies the maximal number of digits in results of formal floating-point operations. All formal floating-point operations yield inequality theorems which approximate real results from above or below. Formal verification procedures are based on our implementation of interval arithmetic which works with formal floating-point numbers. We also cache results of all basic arithmetic operations to improve the performance of formal computations.

A description of our formal verification procedure is technical and it can be found in~\cite{solovyev-thesis}. Here we give an example which demonstrates how the formal verification procedure works. Let $f(x) = x - 2$ and we want to prove $f(x) < 0$ for $x \in [-1, 1]$. Suppose that we have the following solution certificate
\begin{verbatim}
Result_glue {1, false,
   Result_pass_mono {[1, incr]},
   Result_mono {[1, incr],
      Result_pass
   }
}
\end{verbatim}
This certificate tells that the inequality may be verified by first splitting the domain into two subdomains along the first (and the only) variable; then the left branch follows from some other formal verification result by monotonicity (\verb'Result_pass_mono'); the right branch follows by the monotonicity argument and by a direct verification. This certificate cannot be used directly for a formal verification since we don't know how the left branch is proved. The first step is to transform this certificate into a list of certificate such that each certificate can be verified on subdomains specified by the corresponding paths. We get the following list of certificates
\begin{verbatim}
[
  ["r", 1], Result_mono {[1], Result_pass};
  ["l", 1], Result_mono {[1], Result_ref {0}};
  [], Result_glue {1, false, Result_ref {1}, Result_ref {0}}
]
\end{verbatim}
The first element corresponds to the right branch of the original \verb'Result_glue' (hence, the path is \verb'["r", 1]' which means subdivision along the first variable and taking the right subdomain). A formal verification of the first certificate yields $\vdash x \in [0,1] \implies f(x) < 0$. The second result is the transformed left branch of the original certificate. This transformed result explicitly refers to the first proved result (\verb'Result_ref {0}'). Now it can be verified. Indeed, \verb'Result_ref {0}' yields $\vdash x \in [0,0] \implies f(x) < 0$ (since $[0, 0] \subset [0,1]$ and we have the theorem for $[0,1]$ which we use in the reference). Then the monotonicity argument
\begin{align*}
(\forall x,\ x \in [-1,0] \implies 0 \le f'(x)) 
	&\conj (\forall x, x \in [0,0] \implies f(x) < 0) \\
&\implies (\forall x, x \in [-1,0] \implies f(x) < 0)
\end{align*}
yields $\vdash x \in [-1, 0] \implies f(x) < 0$. The last entry of the list refers to two proved results and glues them together in the right order:
\begin{align*}
(\forall x,\ x \in [-1,0] \implies f(x) < 0) 
	&\conj (\forall x,\ x \in [0,1] \implies f(x) < 0)\\
	&\implies (\forall x,\ x \in [-1,1] \implies f(x) < 0)
\end{align*}

% Optimization
\section{Optimization Techniques and Future Work}
\label{nonlinear-optimization}

\subsection{Implemented Optimization Techniques}
There are several optimization techniques for formal verification of nonlinear inequalities. One of the basic ideas of optimization techniques is to compute extra information for solution certificates which helps to increase the performance of formal verification procedures.

The first optimization technique is to try out direct interval evaluations without Taylor approximations. If a direct interval evaluation yields a desired result (verification of an inequality on a domain or verification of a monotonicity property), then a special flag is added to the corresponding certificate node. This flag indicates that it is not necessary to compute full formal Taylor interval and it is enough to evaluate the function directly with interval arithmetic (which is faster). These flags are added to \verb'Result_pass' and \verb'Result_mono' nodes.

An important optimization procedure is to find the best (minimal) precision which is sufficient for verifying an inequality on each subdomain. We have a special informal implementation of all arithmetic, Taylor interval evaluation, and verification functions which compute results in the same way as the corresponding formal functions. This informal implementation is much simpler (because it does not prove anything) and faster (since it does not prove anything and all basic arithmetic is done by native machine arithmetic). For a given solution certificate, we run a modified informal verification procedure which tests different precision parameter values for each certificate node. It finds out the smallest value of the precision parameter for each certificate node such that the verification result is correct. Then a modified solution certificate is created where each node contains information about the best precision parameter. A special version of the formal verification procedure accepts this new certificate and verifies the inequality with computed precision parameters. This adaptive precision technique increases the performance of formal arithmetic computations.

\subsection{Future Work}
There are some optimization ideas which are not implemented yet. The first idea is to stop computations of bounds of second-order partial derivatives for Taylor intervals at some point and reuse bounds computed for larger domains. The error term in Taylor approximation depends quadratically on the size of a domain. When domains are sufficiently small, good approximations of bounds of second-order partial derivatives are not very important. This strategy could save quite a lot of verification time since formal evaluation of second-order partial derivative bounds is expensive for many functions. %Some tests were performed to see how well this approach works with existing solutions certificates. It was shown that about 20\% of evaluations of second-order partial derivative bounds may be eliminated in average. If this strategy is applied to the certificate search procedure, these results could be even better.

Another unimplemented optimization is verification of sets of similar inequalities on the same domain. The idea is to reuse results of formal computations as much as possible for inequalities which have a similar structure and which are verified on the same domains. The basic strategy is to find a subdivision of the domain into subdomains such that each inequality in the set can be completely verified on each subdomain. If inequalities in the set share a lot of similar computations, then the verification of all inequalities in the set could be almost as fast as the verification of the most difficult inequality in the set. This approach should work well for Flyspeck inequalities where many inequalities share the same sub-expressions and domains.

An important unimplemented feature is verification of disjunctions of inequalities. That is, we want to verify inequalities in the form
\[
\forall \bx \in D 
\implies f_1(\bx) < 0 \disj f_2(\bx) < 0 \disj \ldots \disj f_k(\bx) < 0.
\]
This form is equivalent to an inequality on a non-rectangular domain since
\[
(P(\bx) \implies f(\bx) < 0 \disj g(\bx) < 0) \eq (P(\bx) \conj 0 \le g(\bx) \implies f(\bx) < 0).
\]
Many Flyspeck inequalities are in this form. A formal verification of these inequalities is simple. It is enough to add indices of functions for which the inequality is satisfied to the corresponding nodes of solution certificates. Then it will be only necessary to modify the formal gluing procedure. It should be able to combine inequalities for different functions with disjunctions.

% Tests
\section{Results and Tests}
\label{nonlinear-tests}
This section briefly introduces the implemented verification tool and presents some test results for several polynomial and non-polynomial inequalities. We also compare the performance of the formal verification tool and the informal C++ verification procedure for Flyspeck nonlinear inequalities. All tests were performed on Intel Core i5, 2.67GHz running Ubuntu 9.10 inside Virtual Box 4.2.0 on a Windows 7 host; the Ocaml version was 3.09.3; the base of arithmetic was 200.

% Formal Tool
\subsection{Overview of the Formal Verification Tool}
A user manual which contains information about the tool and installation instructions is available at~\cite{nonlinear-manual}. Here, we briefly describe how the tool can be used.

Suppose we want to verify a polynomial inequality
\begin{multline*}
-\frac{1}{\sqrt{3}} \le x \le \sqrt{2} 
	\conj -\sqrt{\pi} \le y \le 1
	\implies x^2 y - x y^4 + y^6 + x^4 - 7 > -7.17995.
\end{multline*}

The following HOL Light script solves this problem
\begin{verbatim}
needs "verifier/m_verifier_main.hl";;
open M_verifier_main;;

let ineq = `-- &1 / sqrt(&3) <= x /\ x <= sqrt(&2)
      /\ -- sqrt(pi) <= y /\ y <= &1
      ==> x pow 2 * y - x * y pow 4 + y pow 6 - &7 + x pow 4 
          > -- #7.17995`;;

let th, stats = verify_ineq default_params 5 ineq;;
\end{verbatim}
First two lines of the script load the verification tool. The main verification function is called \verb'verify_ineq'. It takes 3 arguments. The first argument contains verification options. In most cases, it is enough to provide default options \verb|default_params|. The second parameter specifies the precision of formal floating-point operations. The third parameter is the inequality itself given as a HOL Light term. The format of this term is simple: it is an implication with bounds of variables in the antecedent and an inequality in the consequent. The bounds of all variables should be in the form $\text{\it a constant expression} \le x$ or $x \le \text{\it a constant expression}$. For each variable, upper and lower bounds must be given. The inequality must be a strict inequality ($<$ or $>$). The inequality may include \verb'sqrt' ($\sqrt{}$), \verb|atn| ($\arctan$), and \verb|acs| ($\arccos$) functions. The constant \verb|pi| ($\pi$) is also allowed.

The verification function returns a HOL Light theorem and a record with some verification information which includes verification time.

\subsection{Polynomial Inequalities}

Here is a list of test polynomial inequalities taken from~\cite{MN12}.
\begin{itemize}
% schwefel
\item {\bf schwefel}
\begin{align*}
\langle &x_1, x_2, x_3 \rangle \in [\avec{-10,-10,-10},\avec{10,10,10}]\\
&\implies -5.8806 \times 10^{-10} < 
	(x_1 - x_2^2)^2 + (x_2 - 1)^2 + (x_1 - x_3^2)^2 + (x_3 - 1)^2.
\end{align*}

% caprasse
\item {\bf caprasse}
\begin{align*}
\avec{x_1, x_2, x_3, x_4} &\in [\avec{-0.5,-0.5,-0.5,-0.5},\avec{0.5,0.5,0.5,0.5}]\\
\implies & -3.1801 < -x_1 x_3^3 + 4 x_2 x_3^2 x_4 + 4 x_1 x_3 x_4^2 + 2 x_2 x_4^3 \\
& \phantom{-3.1801 < x} + 4 x_1 x_3 + 4 x_3^2 - 10 x_2 x_4 - 10 x_4^2 + 2.
\end{align*}

% magnetsim
\item {\bf magnetism}
\begin{align*}
\langle x_1,x_2,&x_3,x_4,x_5,x_6,x_7\rangle \in [\avec{-1,-1,-1,-1,-1,-1,-1}, \avec{1,1,1,1,1,1,1}]\\
&\implies
	-0.25001 < x_1^2 + 2 x_2^2 + 2 x_3^2 + 2 x_4^2 + 2 x_5^2 + 2 x_6^2 + 2 x_7^2 - x_1.
\end{align*}

% heart
\item {\bf heart}
\begin{align*}
\langle x_1,x_2,x_3,x_4,x_5,&x_6,x_7,x_8\rangle \in 
	[\avec{-0.1, 0.4, -0.7, -0.7, 0.1, -0.1, -0.3, -1.1},\\
	&\phantom{x_6,x_7,x_8\rangle \in [ }
	\avec{0.4, 1, -0.4, 0.4, 0.2, 0.2, 1.1, -0.3}]\\
\implies &
	-1.7435 < -x_1 x_6^3 + 3 x_1 x_6 x_7^2 - x_3 x_7^3 + 3 x_3 x_7 x_6^2 - x_2 x_5^3 \\
	&\phantom{\text{aaaaaaaaaaa}}
	+ 3 x_2 x_5 x_8^2 - x_4 x_8^3 + 3 x_4 x_8 x_5^2 - 0.9563453.
\end{align*}
\end{itemize}

Performance test results are given in Table~\ref{table-poly}. The column {\it total time} contains total verification time, the column {\it formal} contains time of the formal verification only. The formal verification excludes all preliminary processes: computations of partial derivatives, search of solution certificates, adaptive precision search procedures. The last two columns show the corresponding verification time for the PVS procedure which is based on the Bernstein polynomial technique and described in~\cite{MN12}.

Test results show that our procedure is faster than the Bernstein polynomial procedure in PVS for most cases. On the other hand, there still exist cases where our tool is slower.

\begin{table}[t]
\caption{Polynomial inequalities}
\begin{center}
\begin{tabular}{l@{\quad} r r r r r r}
%{r@{\quad}rl}
\hline
\multicolumn{1}{l}{Inequality ID}&
\multicolumn{1}{l}{\phantom{x}total time (s)}&
\multicolumn{1}{l}{\phantom{x}formal (s)} &
\multicolumn{1}{l}{\phantom{x}total PVS (s)} &
\multicolumn{1}{l}{\phantom{x}formal PVS (s)}\\
\hline\rule{0pt}{12pt}%
schwefel	& 26.33 &	19.15	& 10.23		& 3.18 \\
caprasse	& 8.06	&	1.29 	& 11.44		& 1.25 \\
magnetism	& 7.01 &	1.35	& 160.44	& 82.87 \\
heart		& 17.30 &	1.28	& 79.68		& 26.14 \\
\hline
\end{tabular}
\end{center}
\label{table-poly}
\end{table}

\subsection{Flyspeck Inequalities}
The Flyspeck project contains 985 nonlinear inequalities. The informal verification program written in C++ can verify all these inequalities in about 10 hours. Most inequalities (683) can be informally verified in less than 10 seconds. Almost all inequalities (911) can be informally verified in less than 100 seconds. 

We tested our formal verification procedure on several simple Flyspeck inequalities. Some of these inequalities are listed below. Table~\ref{table-flyspeck} contains performance test results for these inequalities. The column {\it total time} contains total formal verification time, the column {\it formal} contains time of the formal verification only (excluding all preliminary processes), the column {\it informal} contains informal verification time by the C++ program.

\begin{eqnarray*}
\Delta(x_1,\ldots,x_6) &= &x_1 x_4(-x_1 + x_2 + x_3 - x_4 + x_5 + x_6)\\
&& + x_2 x_5(x_1 - x_2 + x_3 + x_4 - x_5 + x_6)\\
&& + x_3 x_6(x_1 + x_2 - x_3 + x_4 + x_5 - x_6)\\
&& - x_2 x_3 x_4 - x_1 x_3 x_5 - x_1 x_2 x_6 - x_4 x_5 x_6,\\[10pt]
\Delta_4 &=& \partd{\Delta}{x_4},
\end{eqnarray*}
\begin{eqnarray*}
\dih_x(x_1,\ldots,x_6) &=& \frac{\pi}{2} - \arctan\left(\frac{-\Delta_4(x_1,\ldots,x_6)}{\sqrt{4 x_1 \Delta(x_1,\ldots,x_6)}}\right),\\[6pt]
\dih_y(y_1,\ldots,y_6) &=& \dih_x(y_1^2, \ldots, y_6^2).
\end{eqnarray*}

\begin{itemize}
% 1
\item {\bf 4717061266}
\begin{align*}
4 \le x_i \le 6.3504 \implies \Delta(x_1, x_2, x_3, x_4, x_5, x_6) > 0.
\end{align*}

% 2
\item {\bf 7067938795}
\begin{align*}
4 \le x_{1,2,3} &\le 6.3504,\ x_4 = 4,\ 3.01^2 \le x_{5,6} \le 3.24^2\\
&\implies \dih_x (x_1, \ldots, x_6) - \pi/2 + 0.46 < 0.
\end{align*}

% 3
\item {\bf 3318775219}
\begin{align*}
2 \le y_i \le 2.52
	\implies 0 < &\dih_y (y_1, \ldots, y_6) - 1.629 - 0.763 (y_4 - 2.52) \\
	&- 0.315 (y_1 - 2.0) + 0.414 (y_2 + y_3 + y_5 + y_6 - 8.0).
\end{align*}

\end{itemize}

% Some Flyspeck inequalities
\begin{table}[t]
\caption{Flyspeck inequalities}
\begin{center}
\begin{tabular}{l@{\quad} r r r r r}
%{r@{\quad}rl}
\hline
\multicolumn{1}{l}{Inequality ID}&
\multicolumn{1}{l}{\phantom{x}total time (s)}&
\multicolumn{1}{l}{\phantom{x}formal (s)}&
\multicolumn{1}{l}{\phantom{x}informal (s)}\\
\hline\rule{0pt}{12pt}%
2485876245a	& 5.530 & 0.058	& 0 \\
4559601669b & 4.679 & 0.048	& 0 \\
4717061266  & 27.1 & 0.250	& 0 \\
5512912661  & 8.860 & 0.086	& 0.002 \\
6096597438a & 0.071 & 0.071	& 0 \\
6843920790  & 2.824 & 0.076	& 0.002 \\
SDCCMGA b   & 9.012 & 0.949	& 0.006 \\
7067938795  & 431   & 387	& 0.070 \\
5490182221  & 1726  & 1533	& 0.375 \\
3318775219  & 17091 & 15226	& 8.000 \\
\hline
\end{tabular}
\end{center}
\label{table-flyspeck}
\end{table}

We also found formal verification time of all Flyspeck inequalities which can be verified in less than one second and which do not contain disjunctions of inequalities. Table~\ref{table-flyspeck-1} summarizes test results. The columns {\it total time} and {\it formal} show total formal verification time and formal verification time without preliminary processes for the corresponding sets of inequalities. The column {\it informal} contains informal verification time for the same sets of inequalities.

Test results show that our formal verification procedure is about 2000--4000 times slower than the informal verification program.

% Flyspeck inequalities which can be verified in 1 second or less
\begin{table}[t]
\caption{Flyspeck inequalities which can be informally verified in 1 second}
\begin{center}
\begin{tabular}{l@{\quad} r r r r}
%{r@{\quad}rl}
\hline
\multicolumn{1}{l}{time interval (ms)}&
\multicolumn{1}{l}{\phantom{x}\# inequalities}&
\multicolumn{1}{l}{\phantom{x}total time (s)}&
\multicolumn{1}{l}{\phantom{x}formal (s)}&
\multicolumn{1}{l}{\phantom{x}informal (s)}\\
\hline\rule{0pt}{12pt}%
0			& 57 &	423		& 2.159	& 0 \\
1--100		& 35 &	5546	& 3854  & 1.134 \\
101--500	& 11 &	12098	& 10451	& 3.944 \\
501--700  	& 14 &	32065	& 28705	& 8.423 \\
701--1000  	& 9  & 	19040	& 16688 & 7.274 \\
\hline
\end{tabular}
\end{center}
\label{table-flyspeck-1}
\end{table}

\bibliographystyle{splncs}
\bibliography{bibliography}

\end{document}